\documentclass[conference]{IEEEtran}
\IEEEoverridecommandlockouts

\usepackage{cite}
\usepackage{amsmath,amssymb,amsfonts}
\usepackage{algorithm}
\usepackage{algorithmic}
\usepackage{graphicx}
\usepackage{textcomp}
\usepackage{xcolor}
\usepackage{url}

\usepackage{subfigure}
\usepackage{booktabs}
\usepackage{multirow}
\usepackage{bbm}

\def\BibTeX{{\rm B\kern-.05em{\sc i\kern-.025em b}\kern-.08em
    T\kern-.1667em\lower.7ex\hbox{E}\kern-.125emX}}
\begin{document}

\title{CE-CoLLM: Efficient and Adaptive Large Language Models Through Cloud-Edge Collaboration}

\author{
  \IEEEauthorblockN{Hongpeng Jin, Yanzhao Wu}
  \IEEEauthorblockA{Florida International University, Miami, FL, USA\\\{hjin008, yawu\}@fiu.edu}
}

\maketitle

\begin{abstract}
Large Language Models (LLMs) exhibit remarkable human-like predictive capabilities. However, it is challenging to deploy LLMs to provide efficient and adaptive inference services at the edge. This paper proposes a novel Cloud-Edge Collaboration framework for LLMs (CE-CoLLM) to tackle these challenges. 
First, we identify the transmission of LLM contextual data between the cloud and edge as a key performance bottleneck, which introduces substantial communication overhead that dominates overall inference latency and makes na\"ive cloud-edge collaboration for LLMs inefficient. 
Second, we introduce a suite of novel techniques, including a latency-aware early exit mechanism and efficient cloud context management, into CE-CoLLM, which collectively reduce communication overhead and preserve LLM inference accuracy. 
Third, we design two adaptive inference modes to accommodate diverse edge environments: (1) a low-latency standalone edge inference mode that enables reliable edge-side independent LLM inference even under unstable network conditions, and (2) a high-accuracy cloud-edge collaborative inference mode that adaptively leverages cloud resources to enhance prediction accuracy. 
Extensive experiments on multiple benchmark datasets demonstrate that CE-CoLLM reduces overall inference time by up to 13.81\% and offloads over 84.53\% of the computational workload from the cloud to the edge, compared to conventional cloud-based LLM deployment, without sacrificing prediction accuracy. 
The code is provided on GitHub at \url{https://github.com/mlsysx/CE-CoLLM}. 
\end{abstract}


\section{Introduction}
Large Language Models (LLMs) have demonstrated remarkable predictive capabilities, transforming diverse fields ranging from Natural Language Processing (NLP) to critical decision-making tasks~\cite{Teubner2023Welcome, naveed2023comprehensive, Chang2024llmsurvey, Badhan2024Security, rangaraj2025effective}. 
There is a growing interest in extending the powerful predictive performance of LLMs directly to edge devices to gain benefits, such as low inference latency, enhanced privacy protection, and reliable inference independent of network connectivity~\cite{Wu2021Parallel,Ganesh2023Amplifying, Hao2024Hybrid, zhang2024edgeshard}.  
However, realizing these objectives presents critical challenges. 
On the one hand, deploying full-scale LLMs directly on resource-constrained edge devices is often impractical due to their limited computational and memory resources~\cite{Wu2021Parallel,Wu2022Comparative,xu2024surveyresourceefficientllmmultimodal,qu2024mobileedgeintelligencelarge}, typically requiring model compression or pruning techniques that compromise model accuracy~\cite{Huang2024Billm,kundu2024efficiently,ma2024era1bitllmslarge,Lin2024awq}. 
On the other hand, relying solely on the conventional cloud-based LLM deployment~\cite{instructgpt,gpt4,anthropic2023claude}, which utilizes substantial computational power in the cloud, introduces inherent performance issues, such as network-dependent communication latency and vulnerabilities under service or network interruptions~\cite{Ganatra2023Detection,Chen2024Automatic,Ding2024Enhancing,Hao2024Hybrid,zhang2024edgeshard}. 
These limitations highlight the pressing need to explore cloud-edge collaboration strategies, which hold the potential to harness the benefits of both cloud and edge computing to deliver efficient, adaptive, and reliable LLM-based services at the edge. 

\begin{figure}[t]
    \centering
    \includegraphics[width=0.48\textwidth]{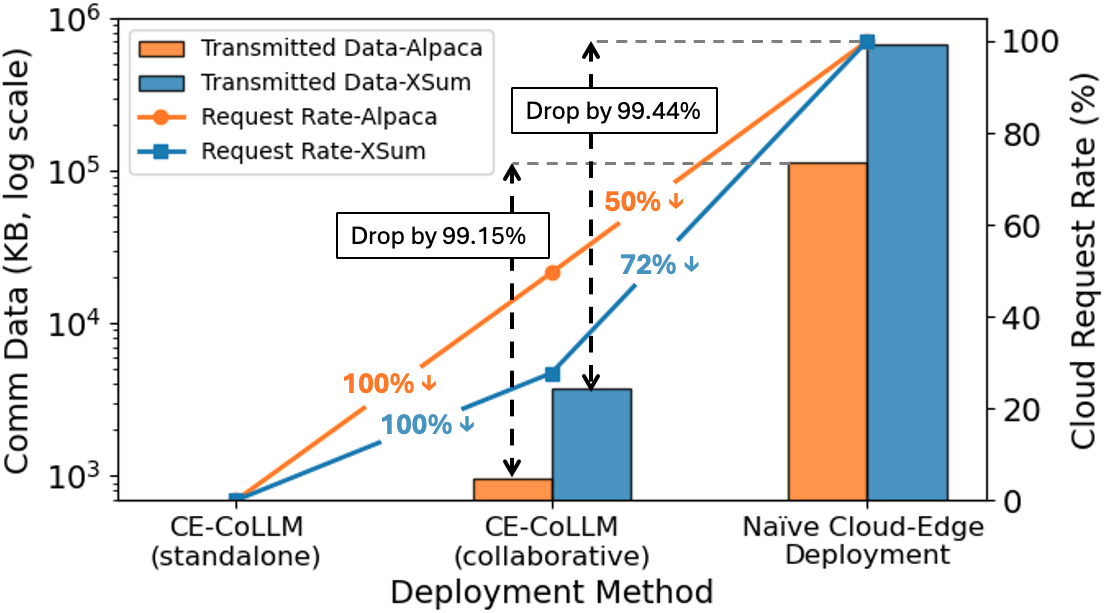}
    \vspace{-3mm}
    \caption{Comparison of average transmitted data sizes per response and cloud request rates between CE-CoLLM and Na\"ive Cloud-Edge Deployment on Alpaca and XSum datasets. Bar plots represent the transmitted data size (log scale, KB), while solid lines indicate the cloud request rate (percentage). The results demonstrate that CE-CoLLM significantly reduces both communication overhead and reliance on cloud requests.}
    \label{fig:comm_plot}
    \vspace{-3mm}
\end{figure}

\begin{figure}[t]
\centering
\subfigure[Cloud LLM Deployment]{
  \centering
  \includegraphics[width=0.225\textwidth]{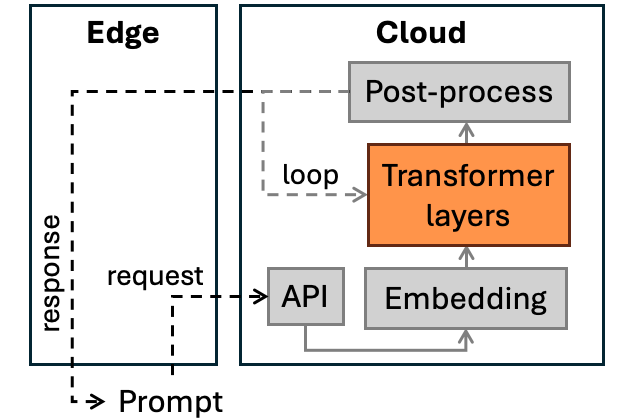}
  \label{fig:cloud_based_llm_deployment}
} 
\subfigure[Na\"ive Cloud-Edge Deployment]{
  \centering
  \includegraphics[width=0.225\textwidth]{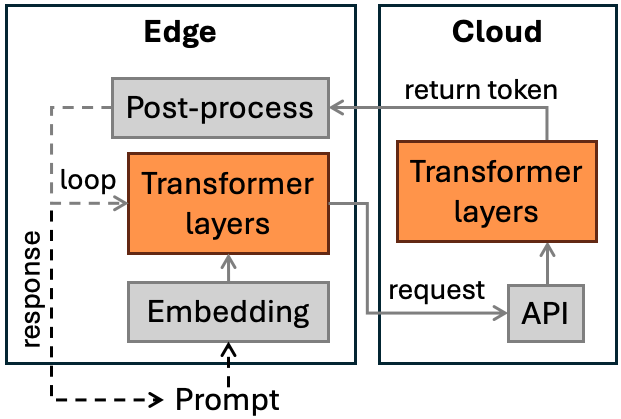}
  \label{fig:naive_edge_cloud_deployment}
} 
\vspace{-4mm} 
\caption{Figure (a) shows the architecture of Cloud LLM Deployment. The edge devices access LLM inference services through
API requests. Figure (b) is the architecture of Na\"ive Cloud-Edge Deployment. In this setup, hidden states are transmitted from the edge device to the cloud for each token inference. The cloud-side computation begins only after receiving these hidden states.}
\label{fig:base_deployment} 
\vspace{-2mm} 
\end{figure}

Several recent studies have made early attempts to enable cloud-edge collaboration for LLM deployment~\cite{Hao2024Hybrid, apple2023intelligence, zhang2024edgeshard}. A popular approach is the hybrid deployment strategy, where a Small Language Model (SLM) is deployed at the edge to handle simple tokens, while complex tokens are offloaded to a powerful LLM in the cloud~\cite{Hao2024Hybrid, apple2023intelligence}. 
While promising, such methods often result in redundant computation or suboptimal resource utilization. A more integrated alternative is to partition a single LLM across the cloud and edge to perform collaborative inference~\cite{zhang2024edgeshard}. However, this split-model deployment introduces substantial communication overhead due to the iterative transmission of a large amount of contextual data (e.g., hidden states) between the cloud and edge for each token, as illustrated in Figure~\ref{fig:naive_edge_cloud_deployment}. 
Our study shows that frequent transmission of contextual data dominates overall inference latency, making the na\"ive cloud-edge collaborative LLM deployments slow and impractical for real-world applications. 

To address these critical challenges, this paper introduces CE-CoLLM, a novel \textbf{C}loud-\textbf{E}dge \textbf{Co}llaborative framework for \textbf{LLM}s. 
Figure~\ref{fig:comm_plot} shows an experimental comparison between CE-CoLLM and na\"ive cloud-edge deployment in terms of the average transmitted data size per response and cloud request rate on the Alpaca~\cite{alpaca} and XSum~\cite{xsum} datasets. CE-CoLLM achieves significant reductions in both communication overhead and cloud request rates, compared to the na\"ive cloud-edge deployment. 
We make three original contributions. 
\textit{First,} we conduct an empirical analysis to show the substantial communication overhead caused by transmitting contextual data in na\"ive cloud-edge deployment. 
\textit{Second,} we propose the CE-CoLLM framework, which integrates a suite of novel components, including the latency-aware early exit mechanism and cloud context management, to effectively mitigate the communication bottleneck. CE-CoLLM offers two flexible inference modes to accommodate diverse edge environments: (1) low-latency standalone edge inference and (2) high-accuracy collaborative cloud-edge inference.  
\textit{Third,} through comprehensive experiments on popular benchmark datasets, we demonstrate that CE-CoLLM significantly outperforms the na\"ive  cloud-edge collaboration by drastically reducing communication overhead, leading to lower end-to-end inference latency while maintaining comparable prediction accuracy. 

\section{LLM Deployment Strategy} \label{sec:cloud_llm_deployment}

In this section, we introduce three main strategies for deploying Large Language Models (LLMs): cloud deployment, edge deployment, and cloud-edge collaborative deployment. 

\noindent \textbf{Cloud LLM Deployment} is a prevalent strategy that enables full-scale Large Language Model (LLM) inference by leveraging ample cloud computational resources. 
Popular LLM-based services, such as ChatGPT~\cite{instructgpt,gpt4} and Claude~\cite{anthropic2023claude}, are hosted in the cloud, providing edge-side users with remote access to these powerful LLMs, as illustrated in Figure~\ref{fig:cloud_based_llm_deployment}. 
However, the cloud LLM deployment is highly dependent on network connectivity and service availability. This dependence incurs significant round-trip communication latency due to the transmission of prompts, responses and intermediate data, which makes the cloud-based LLM service vulnerable to disruptions caused by network instability or cloud-side load surges~\cite{Lin2024awq,Miao2023TowardsEfficientGenerative}, particularly hindering real-time applications at the edge.

\noindent \textbf{Edge LLM Deployment} executes LLMs directly on edge devices (e.g., mobile or IoT devices), bringing inference closer to end users. Representative frameworks include ChatRTX~\cite{nvidia2023chatrtx} and Ollama~\cite{ollama2023website}, which reduce network dependency and enable low-latency services through local inference at the edge. 
However, deploying full-scale LLMs on edge devices is challenging due to their limited compute, memory, and storage resources. To accommodate these constraints, models are often compressed using techniques such as quantization or pruning~\cite{Sun2024Wanda,Dettmers2024SpQR,Frantar2023SparseGPT}. While these methods effectively reduce model size, they can inadvertently compromise prediction accuracy or reduce the model's ability to handle complex tasks~\cite{Huang2024Billm,ma2024era1bitllmslarge,Lin2024awq}, thereby limiting their utility for applications that demand high-quality LLM inference. 

\noindent \textbf{Cloud-Edge Collaborative LLM Deployment} harnesses the extensive computational power of the cloud while utilizing the low-latency advantages of edge devices. 
LLM deployment across cloud and edge environments generally falls into two main categories. 
(1) \textit{Separate Model Deployment:} A full-scale LLM is deployed in the cloud, while a Small Language Model (SLM) runs at the edge. Complex tasks are routed to the cloud to maintain prediction accuracy. However, this strategy can introduce redundant inference overhead if both the edge and cloud process the same input, resulting in suboptimal resource utilization. 
Furthermore, maintaining consistent context and state between the edge and cloud can be challenging~\cite{Hao2024Hybrid,apple2023intelligence}.
(2) \textit{Split Model Deployment:} The LLM is split into edge and cloud partitions, with the initial layers (edge partition) deployed on the edge device and the remaining layers (cloud partition) hosted in the cloud. 
This split model deployment strategy aims to preserve the accuracy of a full-scale LLM. However, due to the auto-regressive nature of LLMs~\cite{vaswani2017attention}, where each token is generated based on preceding tokens, it incurs frequent transmission of substantial intermediate data between the edge and cloud for each generated token. 
As shown in Figure~\ref{fig:comm_plot}, the large volume of transmitted data introduces high communication latency, often dominating the overall inference latency. 

To overcome the communication challenges associated with split model deployment, we propose CE-CoLLM, a novel cloud-edge collaboration framework for LLMs. CE-CoLLM enables adaptive inference task distribution between the cloud and edge, efficiently offloading computational workloads from the cloud to the edge to enhance resource utilization and reduce communication overhead. Furthermore, it provides an edge standalone mode for edge-side users, ensuring resilient and efficient inference at the edge even when the cloud network connection is interrupted.

\section{CE-CoLLM Overview}
\begin{figure}[t]
    \centering
    \includegraphics[width=0.4\textwidth]{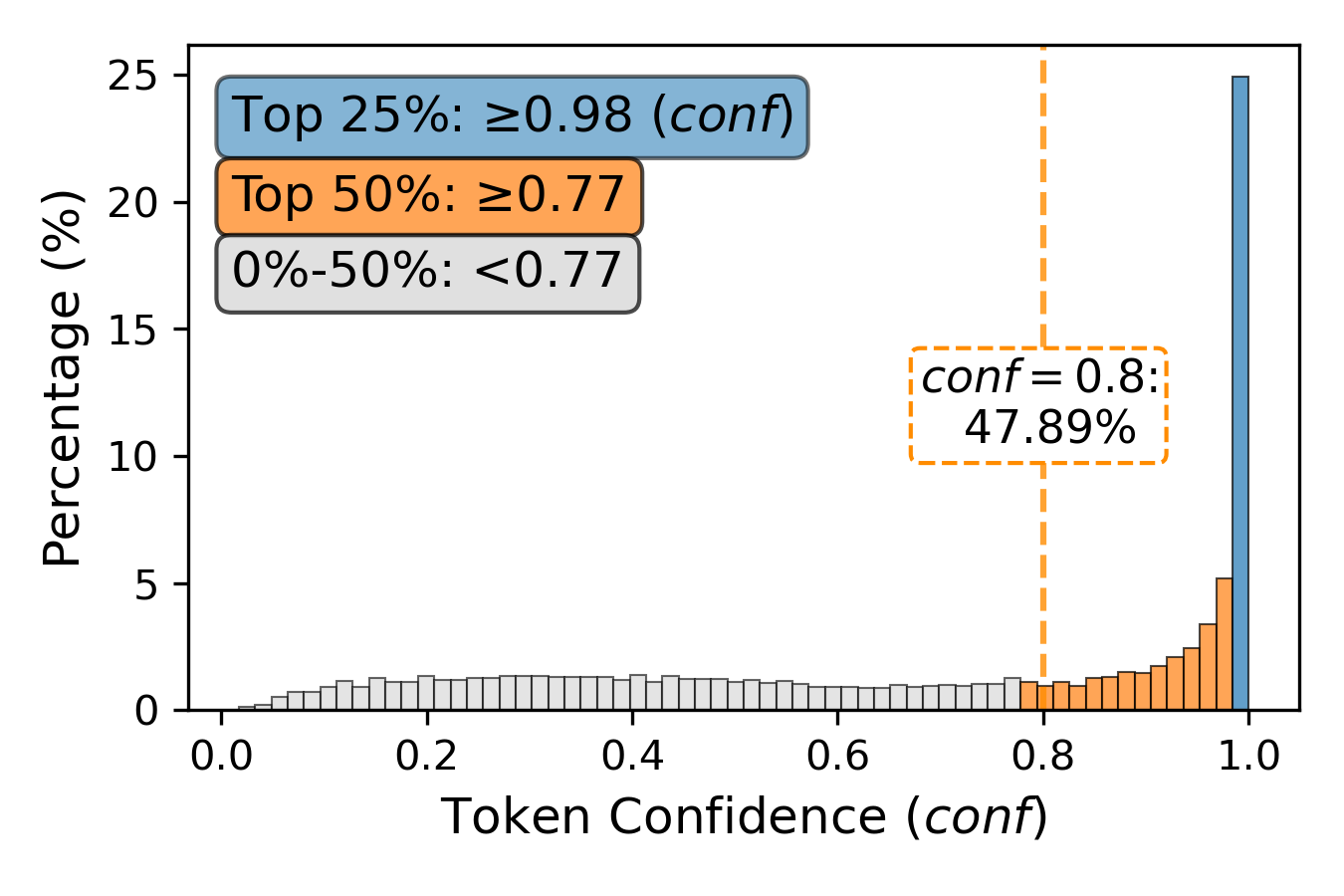}
    \vspace{-5mm}
    \caption{The distribution of token confidence scores at an early exit layer for the Alpaca dataset, showing a long-tail pattern skewed toward high confidence scores, with 50\% of tokens exceeding a confidence score of 0.77 and 25\% exceeding 0.98.}
    \label{fig:token_confidence_plot}
    \vspace{-2mm}
\end{figure}

In this section, we provide an overview of our proposed CE-CoLLM framework, designed to enable efficient and adaptive \textbf{C}loud-\textbf{E}dge \textbf{Co}llaboration for \textbf{LLM}s. 
CE-CoLLM consists of three key functioning components: (1) latency-aware early exit mechanisms, (2) asynchronous contextual data upload, and (3) efficient context management. As illustrated in Figure~\ref{fig:method_overview} (left), CE-CoLLM supports both standalone edge inference and adaptive cloud-edge collaborative inference modes. 
We observed that not all token generations require full LLM inference.  Figure~\ref{fig:token_confidence_plot} shows the distribution of token generation confidence scores, which exhibits a long-tail pattern, with most generated tokens concentrated near the high confidence end. Specifically, 47.89\% of tokens on the Alpaca dataset~\cite{alpaca} and 68.26\% on the XSum dataset~\cite{xsum} achieve confidence scores above 0.8 at an intermediate early exit layer. 
This indicates that a substantial portion of tokens can be generated with high confidence before reaching the final layer of an LLM, presenting opportunities to eliminate unnecessary computation in subsequent layers. 
This observation motivates the core design of CE-CoLLM with a latency-aware early exit mechanism that terminates inference at intermediate layers for high-confidence tokens. Moreover, the early exit mechanism provides a natural way to split an LLM into multiple partitions. The first several layers, along with one or more early exit points, constitute only a small fraction of the LLM and can be deployed on resource-constrained edge devices, while the remaining layers are deployed in the cloud. This design allows the edge device to efficiently process most tokens with high confidence, while more challenging cases (i.e., low-confidence tokens) are offloaded to the cloud to continue the inference, ensuring adaptive and efficient LLM deployment. 
We below describe the key components of CE-CoLLM in facilitating efficient and adaptive cloud-edge collaboration for LLMs. 

\subsection{Latency-aware Early Exit Mechanism}
\begin{figure*}[t]
    \centering
    \includegraphics[width=0.95\textwidth]{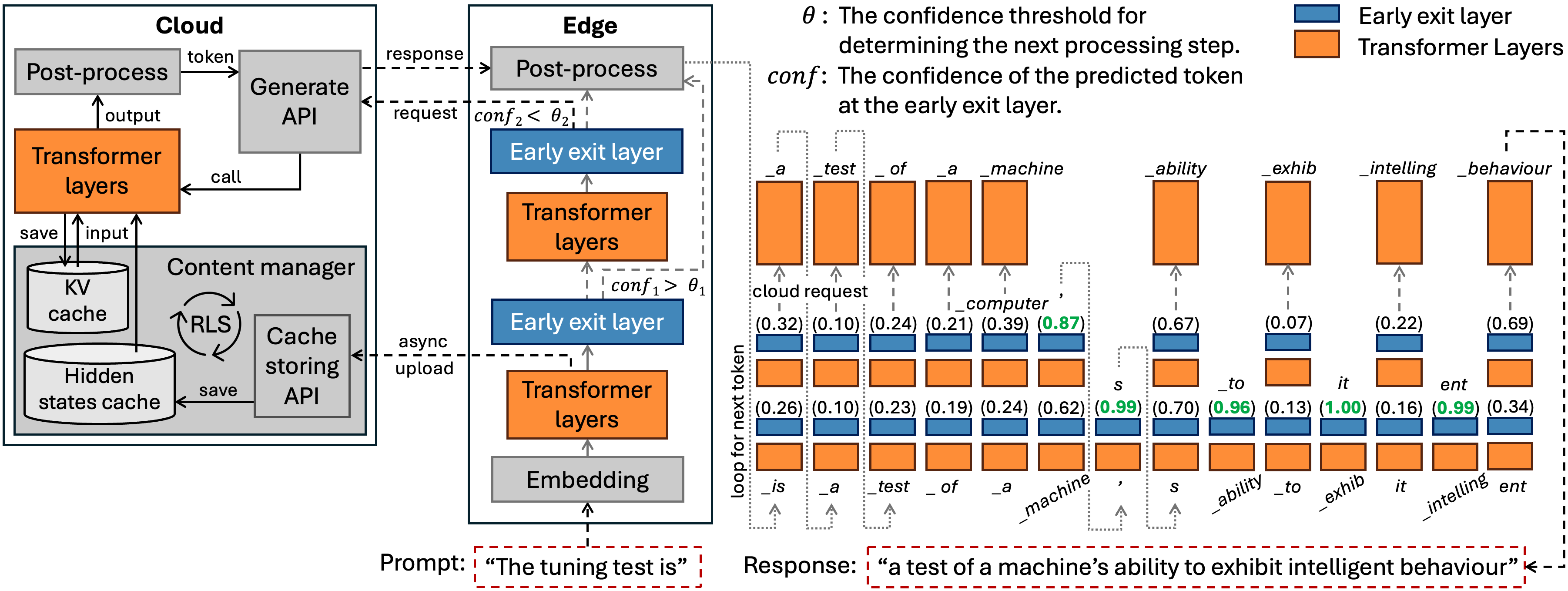}
    \vspace{-3mm}
    \caption{CE-CoLLM architecture overview and workflow: (left) deployment of CE-CoLLM across the edge and cloud with two early-exit points at the edge; (right) end-to-end cloud-edge inference workflow, where tokens with confidence scores $\geq$ 0.8 are generated locally via an early exit point, while these falling below the threshold are offloaded to the cloud for continued inference.}
    \label{fig:method_overview}
    \vspace{-2mm}
\end{figure*}

We incorporate a latency-aware early exit mechanism into CE-CoLLM to optimize the cloud-edge collaboration for LLMs. As shown in Figure~\ref{fig:method_overview} (left), this mechanism integrates multiple early exit points at intermediate layers of the LLM. The initial few layers, equipped with early exit points, constitute a lightweight edge partition suitable for deployment on resource-constrained edge devices, while the remaining layers form the cloud partition for execution in the cloud. 
During inference, each token is first processed through the edge partition and evaluated at each exit points. If the token's confidence score ($conf$) exceeds a predefined threshold ($\theta$), it is generated directly at the early exit point locally without requiring cloud support. The confidence score is computed as:
\[
conf = \max_i \frac{\exp(z_i)}{\sum_{j=1}^V \exp(z_j)},
\] 
where $\{z_1, z_2, \dots, z_V\}$ denote the output logits over a vocabulary of size $V$, and $i, j \in \{1, 2, \dots, V\}$.
If the confidence score at the final edge-side early exit point still falls below the threshold $\theta$, the token is then offloaded to the cloud, where the remaining inference is completed using the cloud partition.

This design enables two adaptive operational modes in CE-CoLLM:
(1) \textit{Edge Standalone Inference Mode (Low-Latency Mode)} is designed for scenarios with limited or unreliable network connectivity. In this mode, the edge-side LLM inference operates independently, generating tokens locally via edge-side early exit points without relying on the cloud. This ensures uninterrupted LLM-based services and low latency responses (see CE-CoLLM (standalone) performance in Table~\ref{tab:single_cost_comparison}). 
(2) \textit{Adaptive Cloud-Edge Collaborative Inference Mode (High-Accuracy Mode)} serves as the default mode under stable network conditions with high prediction accuracy. In this mode, if a token's confidence score remains below the predefined threshold $\theta$ at the edge, CE-CoLLM sends the token along with its necessary contextual information to the cloud for completing the remaining inference. This design enables the edge-side LLM inference to efficiently handle most token generations, while leveraging cloud support for a few challenging tokens to maintain prediction accuracy, as illustrated in Figure~\ref{fig:method_overview} (right). The threshold $\theta$ can be tuned to control the workload distribution between the edge and cloud. By default, we recommend setting $\theta = 0.8$ to achieve a balanced trade-off between inference latency and prediction accuracy. 

\subsection{Asynchronous Contextual Data Upload} 
Figure~\ref{fig:comm_plot} illustrates that transmitting substantial contextual data (e.g., hidden states) when invoking cloud support for low-confidence tokens incurs significant communication overhead. 
To address this issue, CE-CoLLM introduces an asynchronous mechanism for uploading contextual data. 
Since most tokens can be confidently generated at the edge, CE-CoLLM leverages this opportunity to decouple data transmission from synchronous inference process by uploading contextual data in parallel with ongoing edge-side inference. 
Specifically, once edge-side LLM inference reaches a predefined model layer and triggers an upload operation (automatically or upon meeting certain conditions, such as a confidence threshold), it proactively transmits the current contextual data to the cloud-side context manager. 
This design ensures that, if cloud support is ultimately required, the necessary contextual information is already available or nearly transferred, allowing cloud-side inference to begin with minimal data transfer delay. 
By overlapping edge-side computation with data transmission, CE-CoLLM effectively masks a significant portion of the communication overhead, significantly reducing overall inference latency. Furthermore, to further optimize the communication efficiency, we convert the data transmitted from \texttt{float32} to \texttt{float16}, thereby halving the data size without compromising model accuracy, as validated in Section~\ref{sec:precision_impact}.

\subsection{Cloud Support for LLM Inference}
The cloud-side LLM partition compromises the remaining LLM layers not deployed at the edge, which is designed to continue inference for low-confidence tokens offloaded by the edge side. The cloud-side LLM inference incorporates three key mechanisms to support efficient and scalable cloud-edge collaboration. (1) \textit{Context Manager} serves as the core component that manages the contextual data for each edge client. It stores asynchronously uploaded contextual data (e.g., hidden states) from edge clients to ensure that all necessary context is available when cloud-side LLM inference is requested. The context manager also maintains Key-Value (KV) caches generated during cloud-side inference, preserving them across the token generation sequence to avoid redundant computations on previously processed tokens. Unused contextual data and key-value (KV) caches are automatically cleared upon session completion or after a specified period of inactivity, ensuring efficient resource utilization.
(2) \textit{Single-Token Response} is employed by the cloud server to return only one token for each cloud inference request, rather than the entire prediction probability vector. This design effectively reduces communication overhead while still supporting subsequent token generation at the edge. 
(3) \textit{Dual APIs} are provided by the cloud server to support separate execution paths for contextual data uploads and inference processing, allowing both operations to proceed in parallel. 
These mechanisms enable the cloud server to efficiently support edge clients in continuing LLM inference for low-confidence tokens while reducing communication and computation costs. This design supports adaptive, efficient, and scalable cloud-edge collaboration for delivering LLM-based services to multiple edge clients. 

\subsection{Cloud-Edge Collaboration Workflow}
The end-to-end inference workflow of CE-CoLLM is illustrated in Figure~\ref{fig:method_overview} (right, with default $\theta$=0.8).
For each input prompt, inference begins on the edge device, where the edge-side LLM partition processes the input and evaluates the confidence score at each early exit point. Once the confidence score meets or exceeds the threshold ($conf$$\geq \theta$), the token is generated locally without invoking cloud support. 
Meanwhile, CE-CoLLM asynchronously updates contextual data to the cloud in parallel with ongoing edge-side inference. This proactive transmission prepares the cloud server for potential cloud support requests. 
If a token's confidence score remains below the threshold $\theta$ at the final edge-side exit point, the edge client requests cloud support. The cloud-side Context Manager retrieves the uploaded contextual data and stored KV cache to allow the cloud-side LLM partition to resume and complete the remaining LLM inference. The generated token is then returned to the edge client, which appends it to the token sequence and proceeds with generating the next token. This process repeats until the generation is completed. 

\begin{table*}[ht]
\centering
\renewcommand{\arraystretch}{1.2}
\caption{Performance comparison across different LLM deployment strategies}
\resizebox{0.95\textwidth}{!}{%
\begin{tabular}{@{}llrrrr@{}}
\hline
\multicolumn{1}{c}{\textbf{Dataset}} & 
\multicolumn{1}{c}{\textbf{Deployment Strategy}} & 
\multicolumn{1}{c}{\textbf{Total Cost (s)}} & 
\multicolumn{1}{c}{\textbf{Cloud Cost (s)}} & 
\multicolumn{1}{c}{\textbf{Edge Cost (s)}} & 
\multicolumn{1}{r}{\textbf{Comm Cost (s)}} \\ \hline
\multirow{4}{*}{Alpaca} 
& Cloud LLM Deployment & 370.166 $\pm$ 1.692 & 369.799 $\pm$ 1.682 & 0 & 0.367 $\pm$ 0.013 \\ 
& Na\"ive Cloud-Edge Deployment & 3371.761 $\pm$ 34.023 & 253.358 $\pm$ 1.695 & 241.428 $\pm$ 8.832 & 2876.975 $\pm$ 24.815 \\ \cline{2-6}  
& CE-CoLLM (standalone) & \textbf{201.574} $\pm$ 0.278 & 0 & 201.574 $\pm$ 0.278 & 0 \\
& CE-CoLLM (collaborative) & \underline{319.057} $\pm$ 0.438 & 112.723 $\pm$ 0.211 & 192.204 $\pm$ 0.437 & 14.130 $\pm$ 0.181 \\  \hline
\multirow{4}{*}{XSum} 
& Cloud LLM Deployment & 392.466 $\pm$ 7.075 & 392.012 $\pm$ 7.128 & 0 & 0.454 $\pm$ 0.106 \\  
& Na\"ive Cloud-Edge Deployment & 19108.662 $\pm$ 211.580 & 271.661 $\pm$ 5.321 & 411.340 $\pm$ 2.520 & 18425.868 $\pm$ 207.460 \\ \cline{2-6}
& CE-CoLLM (standalone) & \textbf{221.388} $\pm$ 0.865 & 0 & 221.388 $\pm$ 0.865 & 0 \\
& CE-CoLLM (collaborative) & \underline{375.999} $\pm$ 0.720 & 60.652 $\pm$ 0.133 & 258.383 $\pm$ 0.595 & 56.963 $\pm$ 0.137 \\  \hline
\end{tabular}%
} 
\label{tab:single_cost_comparison}
\end{table*}

In summary, CE-CoLLM effectively addresses the communication bottleneck in cloud-edge collaborative LLM deployment through three coordinated mechanisms: (1) latency-aware early exit mechanisms, (2) asynchronous contextual data upload, and (3) efficient context management. This design enables CE-CoLLM to handle most token predictions at the edge, significantly offloading computational load from the cloud to the edge, thereby ensuring efficient resource utilization while maintaining comparable prediction accuracy.

\section{Experimental Analysis} \label{sec:experimental_analysis}  

We conduct experimental analysis primarily from two perspectives to evaluate CE-CoLLM: (1) assessing runtime performance improvements, including faster inference and reduced communication overhead, compared to existing deployment strategies; and (2) verifying that CE-CoLLM preserves prediction accuracy on par with the widely used cloud-based LLM deployment. 
This section focuses on analyzing the performance of four key LLM deployment strategies.
(1) \textit{Cloud LLM Deployment}: A conventional strategy introduced in Section~\ref{sec:cloud_llm_deployment} and illustrated in Figure~\ref{fig:cloud_based_llm_deployment}, serving as our accuracy baseline. 
(2) \textit{Na\"ive Cloud-Edge Deployment}: A baseline strategy that splits an LLM for separate cloud and edge deployments as shown in Figure \ref{fig:naive_edge_cloud_deployment}.
(3) \textit{CE-CoLLM (standalone)}: The standalone mode of our proposed CE-CoLLM framework, operating independently at the edge. 
(4) \textit{CE-CoLLM (collaborative)}: The adaptive cloud-edge collaborative inference mode of our CE-CoLLM framework, using the optimal confidence threshold of $\theta=0.8$.

\noindent \textbf{Experimental Setup:} 
We evaluate CE-CoLLM using a 7B LLaMA model~\cite{llama2} equipped with two early exit points~\cite{Chen2023Eellm}, which is trained to enable adaptive LLM inference by terminating the inference process at an early exit once the confidence exceeds a pre-defined threshold ($\theta$). 
The cloud and edge partitions are deployed separately on a cloud server and an edge device. The cloud server is equipped with a single NVIDIA A100 GPU, which is consistent across all deployment strategies to ensure a fair comparison.
We mainly use two representative datasets, Alpaca~\cite{alpaca} and XSum~\cite{xsum}, to evaluate the LLM inference performance. To validate inference accuracy across different downstream tasks, we employ BoolQ~\cite{boolq} and QuAC~\cite{quac} for question answering, IMDB~\cite{imdb} for sentiment analysis, and XSum for summarization.

\noindent \textbf{Evaluation Metrics:}
We evaluate these LLM deployment strategies using two categories of metrics: (1) \textit{runtime performance metrics}, including total inference time cost, edge computation time cost, cloud computation time cost, and communication time cost, and (2) \textit{accuracy metrics}, including Exact Match (EM)~\cite{rajpurkar2016squad} for BoolQ and IMDB, F1 score for QuAC, and ROUGE-L~\cite{lin2004automatic} for XSum.

\subsection{Runtime Performance Analysis} \label{sec:single_edge} 
We first evaluate runtime performance on a single edge device equipped with an NVIDIA A100 GPU, using 100 randomly selected samples from Alpaca (with relatively short prompts ranging from 13 to 43 tokens) and XSum (with relatively long prompts ranging from 200 to 500 tokens). LLM generates up to 100 tokens for each prompt. The main experiments are repeated five times, and results are reported as mean and standard deviation (mean$\pm$std). Table~\ref{tab:single_cost_comparison} presents the cumulative time cost for 100 prompts, broken down into cloud computation, edge computation, and communication time costs. We highlight three interesting observations.

\textit{First}, the na\"ive cloud-edge deployment suffers from prohibitively high communication costs, making it impractical for real-world applications. Processing a single case takes an average of 33.72 seconds on the Alpaca dataset and 191.09 seconds on the XSum dataset, primarily due to the excessive communication overhead of sending all requests to the cloud. 
In contrast, as illustrated in Figure~\ref{fig:comm_plot}, CE-CoLLM significantly reduces cloud requests to 49.58\% on Alpaca and 27.73\% on XSum, while dramatically decreasing the average data transmitted per request, from 112,128 KB to 956.62 KB on Alpaca (a 99.15\% reduction) and from 673,520.03 KB to 3763.61 KB on XSum (a 99.44\% reduction). 
\textit{Second}, CE-CoLLM significantly outperforms the conventional cloud LLM deployment and delivers enhanced efficiency and inference speed. As shown in Table~\ref{tab:single_cost_comparison}, CE-CoLLM achieves much faster inference of 319 seconds on Alpaca and 376 seconds on XSum, compared to 370 seconds and 392 seconds, for cloud LLM deployment, corresponding to 13.81\% and 4.19\% performance improvements, respectively. 
Furthermore, CE-CoLLM effectively offloads computation from the cloud to the edge, reducing the cloud execution time costs to 113 seconds on Alpaca and 61 seconds on XSum, achieving reductions of 69.52\% and 84.53\%, respectively, compared to cloud LLM deployment. 
This substantial shift of workload to the edge enables the cloud to significantly improve computational efficiency and potentially support more concurrent edge clients using the same resources.
\textit{Third}, CE-CoLLM offers flexible deployment modes to accommodate varying edge environments and network conditions. The standalone mode delivers the fastest inference, with 201.57 seconds on Alpaca and 221.39 seconds on XSum, while completely eliminating cloud dependency, making it suitable for adverse scenarios with unstable network connectivity. Meanwhile, the adaptive cloud-edge collaborative inference mode substantially reduces the cloud computational load by 69.52\% on Alpaca and 84.53\% on XSum, while leveraging the cloud to maintain inference accuracy (see Table~\ref{tab:performance_comparison}). 

\subsection{Impact on LLM Inference Accuracy}
\label{sec:precision_impact}
We then assess the impact of CE-CoLLM on prediction accuracy in comparison to conventional cloud-based LLM deployment across three representative downstream tasks: (1) BoolQ and QuAC for question answering, (2) IMDB for sentiment analysis, and (3) XSum for summarization. Table~\ref{tab:performance_comparison} presents the experimental results. 
Overall, we observe that CE-CoLLM (collaborative) in the cloud-edge collaborative inference mode achieves comparable prediction accuracy to the cloud-based LLM deployment. 
Specifically, for the question answering task, CE-CoLLM delivers an accuracy of 0.658 on BoolQ and 0.289 on QuAC, closely matching the accuracy achieved by cloud LLM deployment of 0.646 and 0.291, respectively. 
For sentiment analysis on IMDB, CE-CoLLM produces the same accuracy as the cloud-based LLM deployment. In the summarization task on XSum, CE-CoLLM achieves an accuracy of 0.225, which is comparable to the cloud LLM deployment with an accuracy of 0.2275. 
These experimental results further validate that CE-CoLLM maintains inference accuracy on par with cloud-based LLM deployment. In the standalone edge inference mode, it incurs only a small accuracy drop while delivering reliable, low-latency inference at the edge without relying on cloud support. 

\begin{table}[t]
\renewcommand{\arraystretch}{1.25}
\centering
\caption{Accuracy comparison between CE-CoLLM (standalone and collaborative modes) and Cloud LLM Deployment}
\resizebox{1.0\linewidth}{!}{
\begin{tabular}{@{}ccccc@{}}
\hline
\multirow{2}{*}{\textbf{Task Type}} & \multirow{2}{*}{\textbf{Dataset}}  & \multicolumn{2}{c}{\textbf{CE-CoLLM}} & \textbf{Cloud LLM} \\
\cline{3-4}
                 &                  & \textbf{Standalone} & \textbf{Collaborative} & \textbf{Deployment} \\ \hline
Question & BoolQ (EM) & 0.4900 & \textbf{0.6580} & 0.6460 \\
Analysis & QuAC (F1) & 0.2740 & 0.2890 & \textbf{0.2910} \\ \hline
Sentiment Analysis & IMDB (EM) & 0.6840 & 0.7240 & 0.7240 \\ \hline
Summarization & XSum (Rouge-L) & 0.1917 & 0.2250 & \textbf{0.2275} \\ \hline
\end{tabular}
}
\vspace{-2mm} 
\label{tab:performance_comparison}
\end{table}

Overall, CE-CoLLM achieves significant performance improvements across multiple dimensions over existing LLM deployment strategies. 
\textit{First,} compared to na\"ive cloud-edge deployment, CE-CoLLM effectively mitigates the communication bottleneck by dramatically reducing both the rate of cloud requests and the volume of transmitted data, making cloud-edge collaboration efficient for real-world applications. 
\textit{Second,} CE-CoLLM significantly outperforms conventional cloud-based LLM deployment in terms of inference speed while effectively offloading computation from the cloud to the edge and preserving inference accuracy, demonstrating the efficacy of the proposed CE-CoLLM framework. 
\textit{Third,} CE-CoLLM provides a standalone edge inference mode that ensures uninterrupted services even under unstable network conditions, thereby enhancing overall system resilience compared to cloud-dependent LLM deployment strategies. 
These advantages position CE-CoLLM as a highly practical solution for delivering efficient and adaptive LLM-based services across diverse edge environments.

\section{Related Work}
The related studies can be summarized into three broad categories: (1) Cloud-Edge Collaborative ML, (2) Early Exit Mechanisms, and (3) Large Language Models. 

\noindent \textbf{Cloud-Edge Collaborative ML} harnesses the high computational power of cloud servers and the low-latency benefits of edge devices to efficiently distribute Machine Learning (ML) workloads across both cloud and edge. While most existing efforts~\cite{Gao2022Cloud,xu2023surveydeepneuralnetwork} focus on deploying conventional Deep Neural Networks (DNNs), cloud-edge collaborative LLM deployment faces unique challenges due to the high communication overhead caused by the inherent iterative inference loops in LLMs. A few recent studies have made early attempts to explore cloud-edge collaboration for LLMs. For instance, \cite{Hao2024Hybrid} deploys a Small Language Model (SLM) at the edge while offloading challenging tokens to a cloud-based LLM, and \cite{zhang2024edgeshard} splits an LLM using dynamic programming to optimize throughput and latency. However, these existing cloud-edge LLM deployment approaches still suffer from high communication overhead and heavy dependence on the cloud, leaving them vulnerable to network disruptions. 

\noindent \textbf{Early Exit Mechanisms} provide an efficient way to dynamically adjust the depth of deep neural network inference, enabling faster and adaptive input processing by allowing the DNN to terminate inference at intermediate layers~\cite{Graves2016Adaptive, Schwartz2020RightTool,Schuster2021Consistent,Ilhan2024Adaptive}. Several studies have incorporated early exit mechanisms into computer vision models~\cite{Teerapittayanon2016Branchynet,Ilhan2024Adaptive} and language models~\cite{Liu2020FastBERT,Hou2020DynaBERT,Zhou2020BERTLosesPatience,Chen2023Eellm}, including recent LLMs~\cite{Corro2023SkipDecode, Bae2023FastAndRobust,Tang2023You,Varshney2024AcceleratingLLaMA,Pan2024Eetuning}. However, few studies investigate efficient frameworks that leverage early exit mechanisms to enable adaptive cloud-edge collaboration for deploying LLMs.

\noindent \textbf{Large Language Models (LLMs)} have achieved remarkable success in various domains~\cite{kaddour2023challenges,ZipZap,shi2025deep}, represented by the GPT series~\cite{gpt1,gpt2,gpt3}, ChatGPT~\cite{instructgpt}, and open-source models, such as LLaMA~\cite{llama,llama2,llama3} and Mistral~\cite{Mistral,Mixtral,Pixtral}. Despite their impressive predictive capabilities, the high computational demands pose critical challenges for delivering LLM-based services at the edge~\cite{Wu2022Comparative,Jin2023Rethinking,Patel2024Splitwise,Zhou2024SurveyEfficientInference}. To tackle these challenges, we propose an efficient and adaptive cloud-edge collaboration framework to empower LLM-based services at the edge. 

\section{Conclusion}
This paper introduces CE-CoLLM, a novel cloud-edge collaboration framework that accelerates LLM inference at the edge with optional cloud support. 
\textit{First,} we identify and quantify the substantial communication overhead in na\"ive cloud-edge LLM deployment, where contextual data transmission dominates inference latency.
\textit{Second,} we propose CE-CoLLM to effectively address this critical communication bottleneck through efficient and adaptive cloud-edge collaboration for LLM deployment.
\textit{Third,} comprehensive experiments on multiple benchmark datasets demonstrate that CE-CoLLM significantly reduces the overall LLM inference latency, cloud computation load, and communication overhead while maintaining comparable prediction accuracy. 
Overall, CE-CoLLM provides an effective and reliable solution for delivering efficient and adaptive LLM-based services at the edge. 

\section*{Acknowledgment}
The authors acknowledge the National Artificial Intelligence Research Resource (NAIRR) Pilot (NAIRR240244), OpenAI, and Amazon Web Services for partially contributing to this research result. 
Any opinions, findings, and conclusions or recommendations expressed in this material are those of the author(s) and do not necessarily reflect the views of funding agencies and companies mentioned above.

{
    \small
    \bibliographystyle{IEEEtran}
    \bibliography{reference}
}

\end{document}